\documentclass[apjl,iop]{emulateapj}

\usepackage{natbib}
\usepackage{graphicx}
\usepackage{txfonts}
\begin{document}

\title{Determination of Transverse Density Structuring from Propagating\\ MHD Waves in the Solar Atmosphere}

\shorttitle{Determination of Coronal Transverse Density Structuring}

\shortauthors{Arregui, Asensio Ramos, \& Pascoe}

\author{I. Arregui\altaffilmark{1,2}, A. Asensio Ramos\altaffilmark{1,2},  and D. J. Pascoe\altaffilmark{3} }
\altaffiltext{1}{Instituto de Astrof\'{\i}sica de Canarias, V\'{\i}a L\'actea s/n, E-38205 La Laguna, Tenerife, Spain}
\altaffiltext{2}{Departamento de Astrof\'{\i}sica, Universidad de La Laguna, E-38205 La Laguna, Tenerife, Spain}
\altaffiltext{3}{School of Mathematics and Statistics, University of St. Andrews,  St. Andrews, KY16 9SS, UK}

 \email{iarregui@iac.es}

\begin{abstract}
We present a Bayesian seismology inversion technique for propagating magnetohydrodynamic (MHD) transverse waves observed in coronal waveguides. The technique uses theoretical predictions for the spatial damping of propagating kink waves in transversely inhomogeneous coronal waveguides. It combines wave amplitude damping length scales along the waveguide with theoretical results for resonantly damped propagating kink waves to infer the plasma density variation across the oscillating structures. Provided the spatial dependence of the velocity amplitude  along the propagation direction is measured and the existence of two different damping regimes is identified, the technique would enable us to fully constrain the transverse density structuring, providing estimates for the density contrast and its transverse inhomogeneity length scale.
\end{abstract}

 
\section{Introduction}

Observations show that propagating magnetohydrodynamic (MHD) waves are present in the solar atmosphere. 
Small amplitude propagating transverse MHD waves have received particular attention, because of their potential for energy storage, transfer, and deposition in the context of wave-based plasma heating processes. A key problem yet to be solved is the quantification of the role of waves in coronal heating. The solution requires a reliable knowledge of the physical properties in magnetic and plasma structures. Because physical properties, such as the magnetic field strength, the plasma density, and their field-aligned and cross-field structuring  cannot be directly measured, seismology of MHD waves offers an alternative method to probe coronal plasmas. 

The presence of ubiquitous coronal transverse waves has been reported by \citet{tomczyk07}  and by \citet{tomczyk09} in observations  taken with the Coronal Multi-Channel Polarimeter (CoMP). The disturbances have amplitudes of the order of 0.3 km s$^{-1}$ and propagate at speeds of  about 0.6 megameters per second. They are interpreted as transverse MHD waves of Alfvenic character.  The measured discrepancy between inward and outward power associated to the disturbances \citep{tomczyk09} is thought to be an indication of in situ wave damping. Resonant absorption of kink waves has been proposed to explain this damping by \citet{pascoe10}. This process, studied in the context of standing kink waves in coronal loops \citep[see e.g.][]{goossens92a,ruderman02,goossens02a,goossens06} and in prominence fine structures \citep{arregui08a,soler09c} occurs because of transverse inhomogeneity of the plasma across the waveguides. Resonant absorption predicts a selective damping as a function of frequency \citep{terradas10}. The good agreement between the observationally measured and the theoretically predicted  outward/inward power ratio as a function of wave frequency strongly supports the idea that resonant absorption operates on these waves \citep{verth10}.

The use of wave damping properties to perform seismology inversions with transverse waves was proposed by \citet{arregui07a} and \citet{goossens08a} in the context of standing kink waves in coronal loops, and by \cite{arregui11a} in the context of prominence fine structures \citep[see reviews by][]{goossens08b, arregui12a,arregui12b}. Because the number of unknowns is usually larger than that of observables, density measurements need to be used as additional observational information to fully constrain the parameters of interest \citep{arregui11b}. 

The mechanism of resonant absorption does not make a distinction between standing and propagating MHD kink waves. The dynamics corresponds to a surface Alfv\'en wave damped by resonant absorption \citep{goossens09,goossens12a}. For standing kink waves, the observational consequence is  the attenuation of the amplitude in time. For propagating kink waves, the attenuation of wave amplitude in space.

\cite{goossens12b} have presented a seismology inversion scheme for propagating MHD waves damped by resonant absorption.  In their analysis, a solution curve that relates the Alfv\'en speed, the density contrast and the transverse inhomogeneity length scale is obtained.  The Alfv\'en speed is constrained to a narrow range, but the two parameters that define the cross-field density structuring cannot be inferred.

\cite{pascoe12} have pointed out that a more general damping profile consisting of a Gaussian and an exponential profile better reproduces the spatial decay of the amplitude for propagating kink waves. A Gaussian seems to better reproduce the amplitude behavior initially, while the exponential profile properly describes the dynamics after a few wavelengths. Their numerical solutions point to a transition between the two damping regimes.  Currently, only indirect observational evidence about in situ damping of propagating transverse waves is available, with no measurement of the damping length-scales. The confirmation on the existence of two damping regimes would make accessible information on two damping length scales and the height at which the transition between the two regimes occurs.

In this Letter,  we demonstrate how this information can be used to fully constrain the cross-field density structuring in coronal waveguides. 

\section{Spatial Damping of Propagating Kink Waves}

The theory for the spatial damping of propagating kink waves due to resonant absorption has been developed by a number of studies \citep{terradas10, verth10,soler11a,soler11c,soler11d,hood13}. 
Numerical simulations have confirmed the obtained damping properties, by analyzing spatial and temporal properties of the mode coupling process in coronal loops and arbitrary inhomogeneous coronal structures \citep{pascoe10,pascoe11,pascoe12,pascoe13}.  The classic theoretical model assumes that transverse kink waves are guided by plasma structures.  The waveguides are considered to be density tubes in the zero plasma-$\beta$ approximation with a uniform magnetic field directed along the axis of a straight cylindrically symmetric structure. They have a uniform internal density, $\rho_i$, and a uniform external density, $\rho_e$, with $\rho_i>\rho_e$, connected in the transverse direction by a non-uniform region of thickness $l/R$, with $R$ the tube radius. Because of the non-uniformity, resonant absorption produces the decay of the wave amplitude in space as the wave propagates. 

For propagating waves with a real frequency $\omega$, an analytical expression for the wavelength in the thin tube approximation is \citep[see][]{terradas10}

\begin{equation}
\lambda=\sqrt{2} v_{\rm A,i} T \left(\frac{\zeta+1}{\zeta}\right)^{-1/2}.
\end{equation}
Here $v_{\rm A,i}$ is the internal Alfv\'en velocity, $T$ the period, and 
$\zeta=\rho_{\rm i}/\rho_{\rm e}$ the density contrast. This solution is valid provided $\omega R/v_{\rm A,i}<< 1$. It expresses a relation between two observable quantities (wavelength and period) and  two quantities to be determined (Alfv\'en velocity and density contrast).

Because of resonant absorption, spatial damping occurs, and the transverse velocity amplitude decays with an exponential profile of the form $\exp(-z/L_{\rm d})$. Under the thin tube and thin boundary ($l/R<<1$)  approximation,  an expression for the damping length, $L_{\rm d}$, as a function of the relevant physical parameters can be obtained. In units of the wavelength this expression is \citep[see][]{terradas10}

\begin{equation}\label{expdamp}
\frac{L_{\rm d}}{\lambda}=  \left(\frac{2}{\pi}\right)^2\left(\frac{R}{l}\right)\left(\frac{\zeta+1}{\zeta-1}\right).
\end{equation}
The first factor is due to the assumed linear density profile at the non-uniform layer. 
Note that the right hand-side of this expression is identical to the one for the damping time over the period
for standing kink waves. The reason is that resonant absorption does not make any 
distinction with respect to the standing or propagating character of the wave.

\begin{figure}
\epsscale{1.00}
\plotone{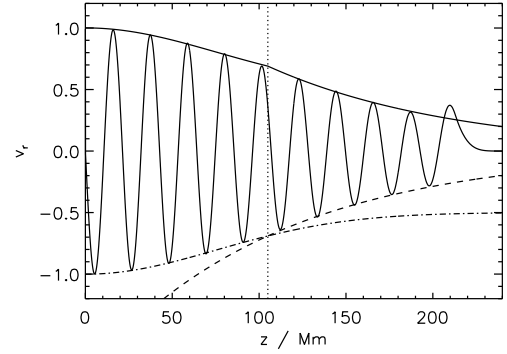}
\caption{Transverse velocity component as a function of height at the axis of the tube for propagating kink waves for a numerical simulation with $\zeta=1.5$ and $l/R=0.4$ (case 4 in Table~\ref{table:numerical}).  On the top is the general spatial damping profile given by the solid line. The transition between Gaussian and exponential damping is given by the vertical dotted line. On the bottom, the general profile is split into its two components; Gaussian (dot-dash) and exponential (dashed).}\label{dampingprofile}
\end{figure}

The exponential profile obtained for standing \citep[e.g.,][]{ruderman02,goossens02a} and propagating \citep[e.g.,][]{terradas10} kink waves describes the asymptotic state of the damping behavior, i.e. at large times or distances. \cite{pascoe12} demonstrated with numerical simulations that the initial damping stage can be described by a Gaussian profile of the form $\exp(-z^2/L^2_{\rm g})$, with $L_{\rm g}$ the Gaussian  damping length scale. \cite{hood13} considered the problem analytically and produced an expression for the full nonlinear spatial damping profile, which can be approximated as Gaussian for low heights and exponential at large heights. Instead, \cite{pascoe13} proposed a general spatial damping profile composed of a Gaussian damping profile at low heights and the usual exponential profile at large heights. An example of the spatial dependence of the velocity amplitude from numerical simulations and the double profile fitting for such a general damping profile is displayed in Figure~\ref{dampingprofile}. The accuracy of this approximate damping profile was demonstrated by the  parametric study performed by \cite{pascoe13}. This study shows that the Gaussian damping length scale can be well described by the expression

\begin{equation}\label{gaussdamp}
\frac{L_{\rm g}}{\lambda}= \left(\frac{2}{\pi}\right) \left(\frac{R}{l}\right)^{1/2}\left(\frac{\zeta+1}{\zeta-1}\right).
\end{equation}
This equation  expresses the Gaussian damping length as a function of the same two parameters that determine the exponential damping length. This means that the observational identification of two damping regimes and the measurement of their associated length scales would provide us with additional information without the inclusion of new model parameters.  The height, $h$, at which the damping regime changes from Gaussian to exponential is given by {\citep[see][]{pascoe13}

\begin{equation}\label{ratio}
h=\frac{L^2_{\rm g}}{L_{\rm d}}=\lambda\left(\frac{\zeta+1}{\zeta-1}\right).
\end{equation}

The non-linear (non-exponential) damping rate is an inherent property of resonantly damped kink oscillations, as derived analytically by \cite{hood13} for spatial damping and by \cite{ruderman13} for time damping. The previous analyses by e.g., \cite{ruderman02} and \cite{goossens02a}, only attempt to describe the damping in the asymptotic state and so they only obtain the later exponential damping stage. 

The observational identification of this feature would be strong support for the resonant damping mechanism. However, no measurement of the axial damping spatial scales is available yet.  Our analysis shows that, if present,  Equations~(\ref{expdamp})--(\ref{ratio}) can be used to fully determine the transverse density structuring in oscillating coronal waveguides.} 

Out of the three equations, only two are independent. In our inversion scheme, Eqs.~(\ref{gaussdamp}) and (\ref{ratio}) will be used to infer $\zeta$ and $l/R$ from data given by  $L_{\rm g}$ and $h$.   The two algebraic equations can be directly solved for the two unknowns.  However, the problem we intend to solve is not exact, since real data are noisy and measurements have an associated uncertainty. 
To accommodate a proper propagation of uncertainty,  we make use of Bayesian analysis, a 
framework that is starting now to be developed in coronal seismology \citep{arregui11b,arregui13}.

\section{Bayesian Inversion Scheme}

To perform our inversion we use Bayes' theorem \citep{bayes73}

\begin{equation}
p(\mbox{{\boldmath $\theta$}} | d)=\frac{p(d | \mbox{{\boldmath $\theta$}})p(\mbox{{\boldmath $\theta$}})}{p(d)},\label{bayes}
\end{equation}
which gives the solution to the inverse problem in terms of the posterior probability distribution, $p(\mbox{{\boldmath $\theta$}} | d)$, that describes how probability is distributed among the possible values of the unknown parameter, {\boldmath $\theta$}, given the data $d$.  The posterior is a combination of what is known about the parameters before taking into account the data, the prior $p(\mbox{{\boldmath $\theta$}})$, and the likelihood function, 
$p(d |\mbox{{\boldmath $\theta$}})$, which tells us how well the model predicts the observed data. The denominator, $p(d)$, is the so-called evidence, which normalises the likelihood. It plays no role in parameter inference.

Once the posterior is known, we calculate how a particular parameter is constrained by data computing its marginal posterior. This is done by performing the following integral of the full posterior with respect to the rest of parameters

\begin{equation}\label{marginals}
p(\theta_i|d) = \int p(\mbox{{\boldmath $\theta$}} | d) d\theta_1 \ldots
d\theta_{i-1} d\theta_{i+1} \ldots d\theta_{N}.
\end{equation}
\noindent
The result provides us with all the information for model parameter $\theta_i$ available in  the priors and the data. This method also enables us to correctly propagate uncertainties from data to inferred parameters.

We next specify the likelihood function and the priors.  In what follows we assume the observed data are given by  $d=(L_{\rm g}$, $h$), where both observed length-scales are normalized to the wavelength. The unknowns are gathered in the vector {\boldmath $\theta$}=($\zeta$, $l/R$). Under the assumption that observations are corrupted with Gaussian noise and they are statistically independent, the likelihood can be expressed as

\begin{equation}\label{likelihood}
p(d|\mbox{\boldmath $\theta$}) =\left(2\pi \sigma_{L_{\rm g}} \sigma_{h}\right)^{-1} \exp \left\{
\frac{\left[L_{\rm g}-L^\mathrm{syn}_{\rm g}(\mbox{\boldmath $\theta$})\right]^2}{2 \sigma_{L_{\rm g}}^2} 
+ \frac{\left[h-h^\mathrm{syn}(\mbox{\boldmath $\theta$})\right]^2}{2 \sigma_{h}^2}
\right\},
\end{equation}

\noindent
with $L_{\rm g}^\mathrm{syn}(\mbox{\boldmath $\theta$})$ and $h^\mathrm{syn}(\mbox{\boldmath $\theta$})$  given by Equations~(\ref{gaussdamp}) and (\ref{ratio}). Likewise, $\sigma_{L_{\rm g}}^2$ and $\sigma_{h}^2$ are the variances associated to the Gaussian damping length and the height $h$, respectively.

The priors indicate our level of knowledge (ignorance) before considering the observed data. We have adopted uniform prior distributions for both unknowns over given ranges, so that we can write

\begin{equation}
p(\theta_i)=\frac{1}{\theta^{\mathrm{max}}_i-\theta^{\mathrm{min}}_i} \mbox{\hspace{0.2cm}}\mbox{for}
\mbox{\hspace{0.2cm}}\theta^{\mathrm{min}}_i\leq\theta\leq\theta^{\mathrm{max}}_i,
\end{equation}
and zero otherwise. For the minimum and maximum values the intervals  $\zeta\in[1,20]$ and $l/R\in[0,2]$ 
have been taken. This choice of priors expresses our belief that the unknown parameters are constrained to those ranges, with all combinations being equally probable. 

Computations were repeated for different uniform prior ranges for $\zeta$, with $\zeta_{\mathrm{min}}=1$ and 
 $\zeta_{\mathrm{max}}=5, 10, 50$, and for $l/R$, with $(l/R)_{\mathrm{min}}=0$ and $(l/R)_{\mathrm{max}}=1, 1.5, 2$.
 We also used strongly informative priors (Gaussians in $\zeta$ and $l/R$ centred far away the synthetic parameter values). For strong prior information on $l/R$, convergence to the correct posterior is achieved through repeated application of Bayes' theorem, using the posterior of one iteration as a prior for the next one and multiplying again by the likelihood. This means that a few consistent observations of the data lead to a converged  posterior, independently of the prior.  Our sensitivity analyses lead us to conclude that posteriors are dominated by the information contained in the data, that overwhelms the prior information.

Posteriors are evaluated for different combinations of parameters using Bayes' theorem (\ref{bayes}) and inferred distributions are obtained from the computation of the marginal posteriors using Eq.~(\ref{marginals}).
To solve the 1D integrals, numerical quadratures were employed, using an adaptive Gauss-Kronrod quadrature.  All computations were checked by comparison to results obtained with the MCMC code employed by \cite{arregui11b}.

\begin{figure*}[t]
   \centering
   \includegraphics[angle=0,scale=0.5]{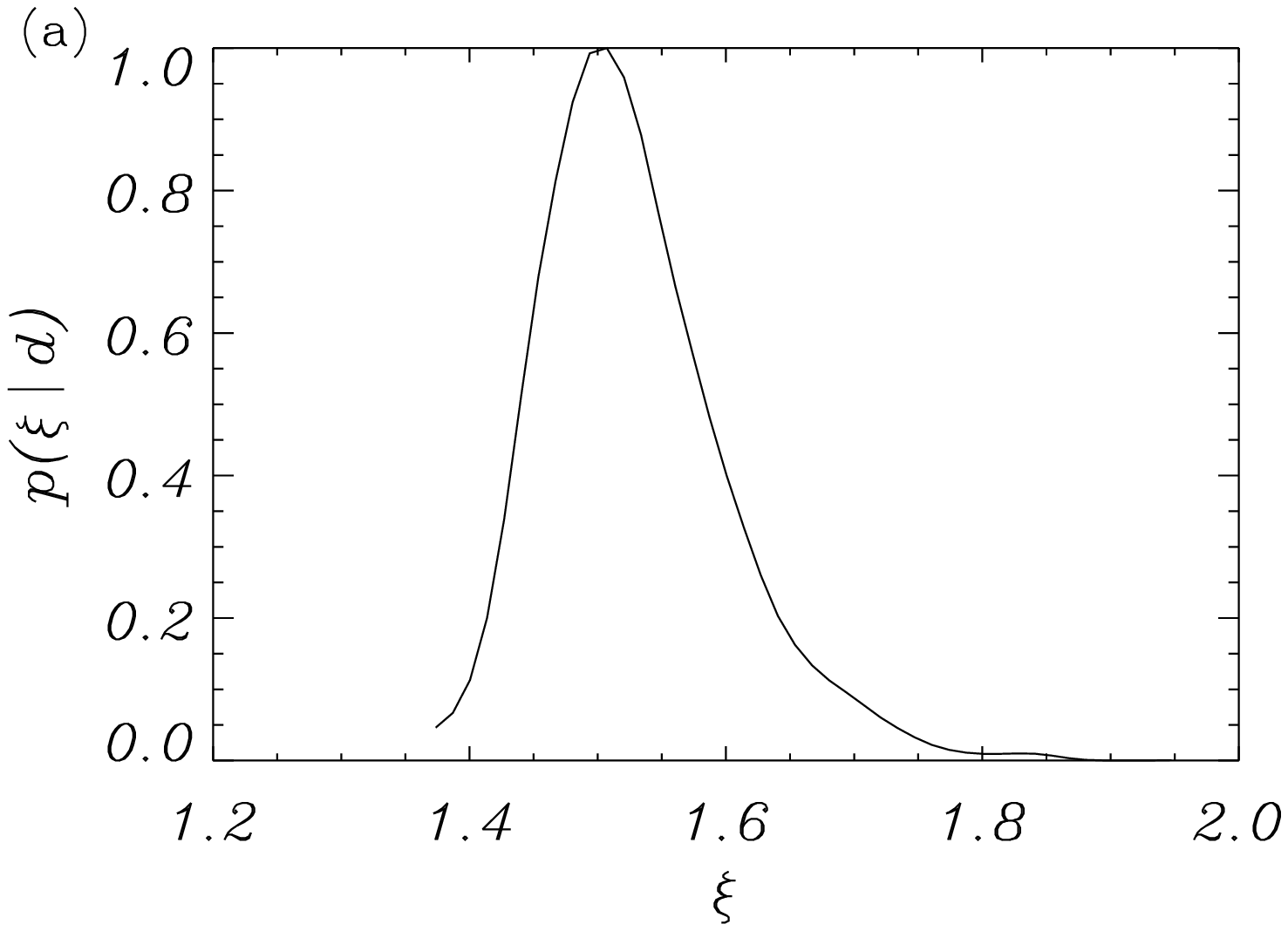} 
     \includegraphics[angle=0,scale=0.5]{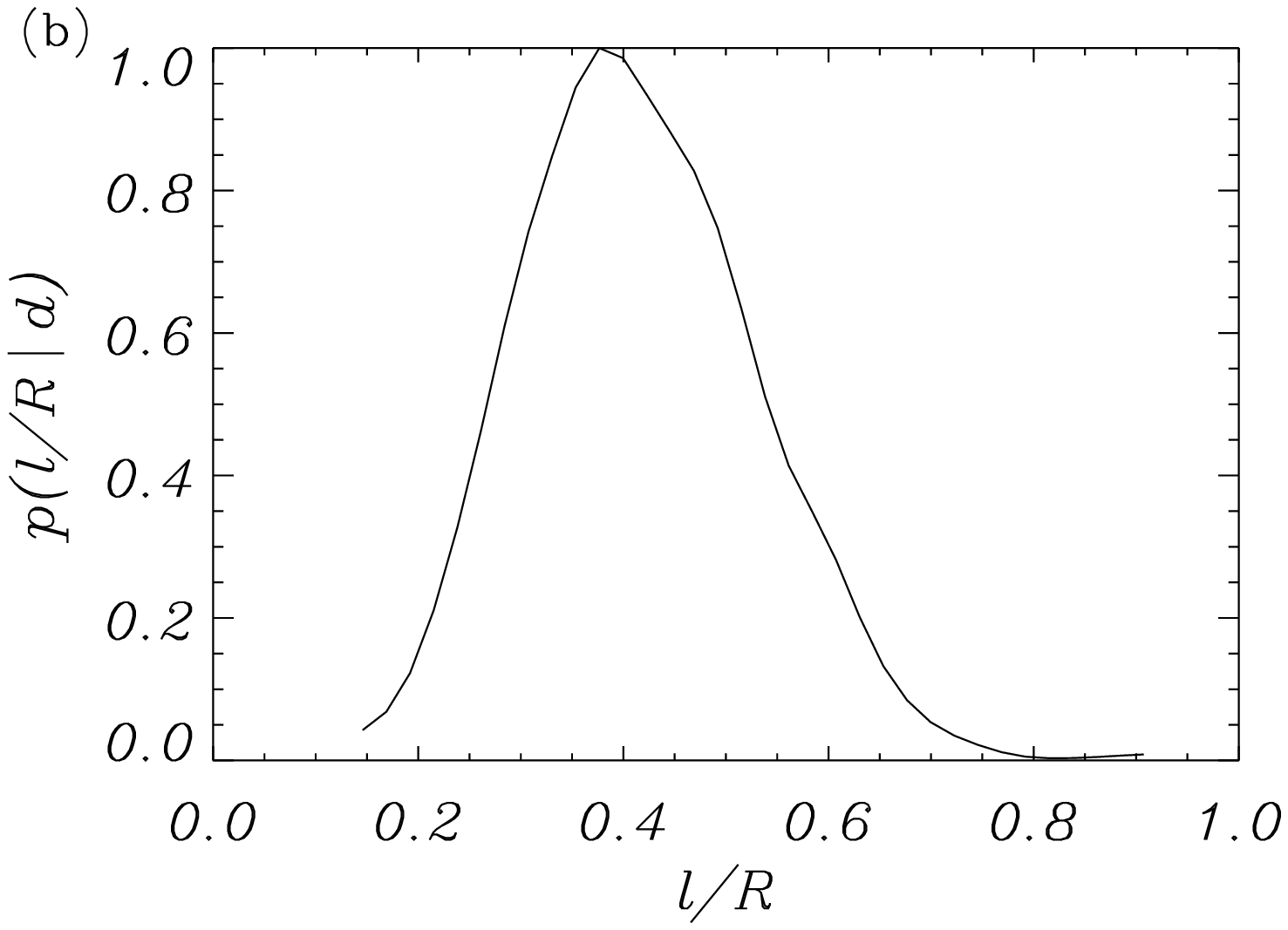} 
   \includegraphics[angle=0,scale=0.5]{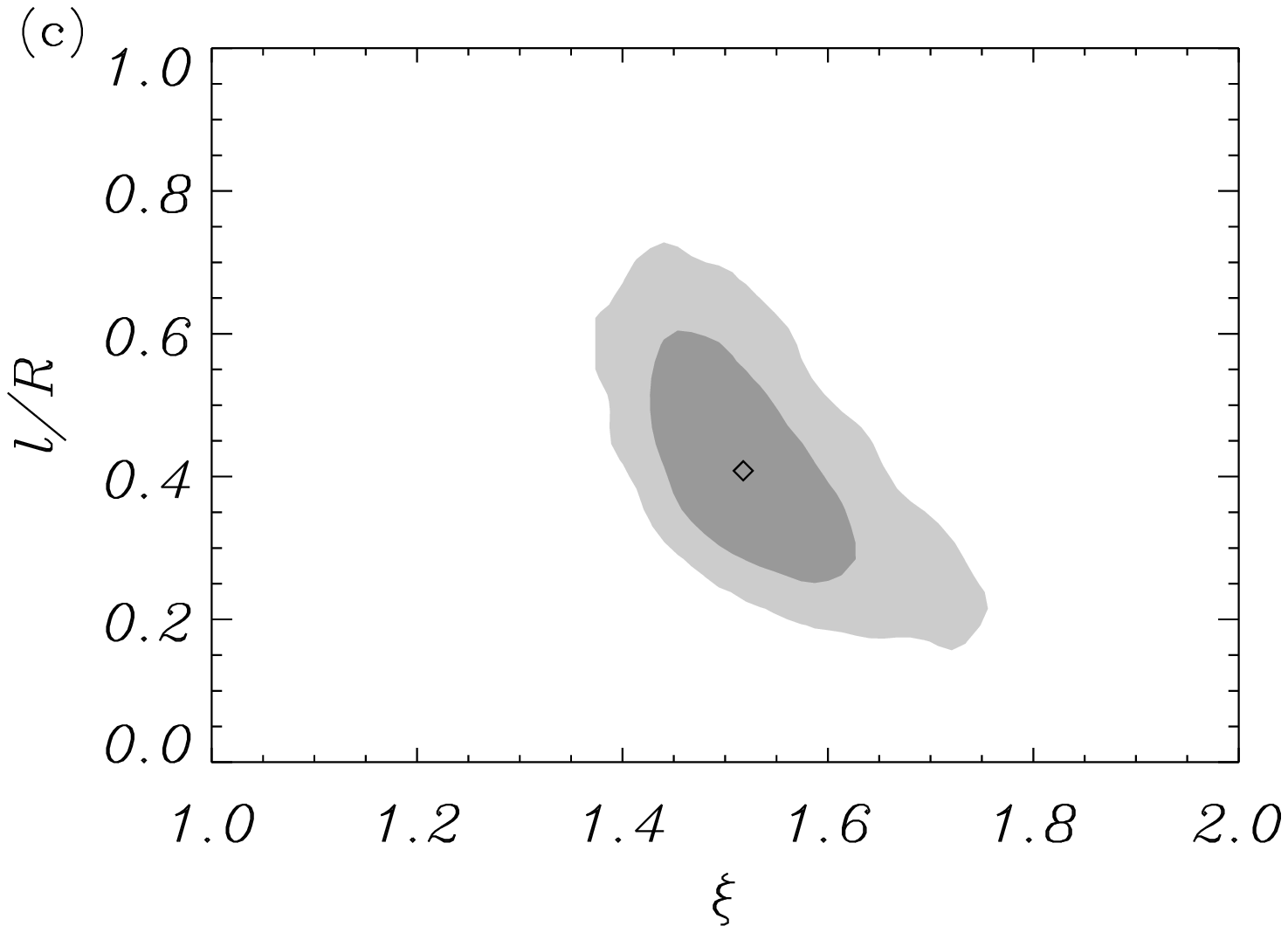} 
    \caption{One-dimensional marginalized posterior distributions for the density contrast (a) and the transverse inhomogeneity length scale (b) corresponding to the inversion of a spatially damped transverse oscillation with $L_{\rm g}/\lambda=5.0\pm 0.1$ and $h/\lambda=4.9\pm0.1$ (uncertainties of 10\% have been used). (c) Joint two-dimensional posterior distribution. The light and dark grey shaded regions cover the 95\% and 68\% credible regions. The symbol indicates the estimate.}
   \label{fig:posteriors}
\end{figure*}

\section{Inversion Results}

We first evaluated the performance of our inversion scheme by making the inference under controlled conditions. We generated predictions for the length scales $L_{\rm g}$ and $h$ for different combinations of the equilibrium parameters, $\zeta=1.5, 2, 3, 4$ and $l/R=0.05, 0.15, 0.2, 0.4$, using Equations~(\ref{gaussdamp}) and (\ref{ratio}). Those synthetic data where treated as observed data in the Bayesian inversion. A 10\% uncertainty on the data was considered and the posterior distributions for $\zeta$ and $l/R$ were computed, using the likelihood function~(\ref{likelihood}) and the uniform priors.
Once the posteriors were known, the median and the variances associated to the 68\% confidence level were calculated.  Table~\ref{table:synthetic} displays the inversion results for some parameter combinations. In all the cases, 
the inversion method correctly recovers the values for the physical parameters. The larger the density contrast, the shorter the two length-scales $L_{\rm g}$ and $h$ are. This increases the errors in the inferred density contrast, while errors in $l/R$ are not affected that much. For the combinations with the largest $\zeta=10$ and $l/R=0.5,1,1.5$, $L_{\rm g}$ and $h$ are comparable to the wavelength. This would make very problematic the observational identification of the two damping regimes.


\begin{deluxetable}{cccccccc} 
\tablecolumns{8} 
\tablewidth{0pc} 
\tablecaption{Inversion of Synthetic Data Using the Analytical Forward Model \label{table:synthetic}} 
\tablehead{ 
\multicolumn{2}{c}{Synthetic Parameters}  & \colhead{}& \multicolumn{2}{c}{Synthetic Data}  &\colhead{} & \multicolumn{2}{c}{Inversion Results} \\ 
\cline{1-2}\cline{4-5}\cline{7-8} \\ 
\colhead{$\zeta$} & \colhead{$l/R$} & \colhead{} & \colhead{$L_{\rm g}/\lambda$}  &\colhead{$h/\lambda$}    & \colhead{} & \colhead{$\zeta$}    & \colhead{$l/R$}}
\startdata 
1.5 & 0.05 && 14.2 & 5.0 && 1.51$^{+0.08}_{-0.06}$ & 0.05$^{+0.02}_{-0.01}$\\
1.5 & 0.15 && 8.2 & 5.0 && 1.50$^{+0.07}_{-0.06}$ & 0.16$^{+0.05}_{-0.04}$\\
1.5 & 0.2 && 7.1 & 5.0 && 1.51$^{+0.07}_{-0.06}$ & 0.21$^{+0.06}_{-0.05}$\\
1.5 & 0.4 && 5.0 & 5.0 && 1.50$^{+0.07}_{-0.05}$ & 0.44$^{+0.13}_{-0.11}$\\
3 & 0.05 && 5.7 & 2.0 && 3.11$^{+0.59}_{-0.38}$ & 0.05$^{+0.02}_{-0.01}$\\
3 & 0.15 && 3.3 & 2.0 && 3.09$^{+0.61}_{-0.40}$ & 0.15$^{+0.05}_{-0.04}$\\
3 & 0.2 && 2.9 & 2.0 && 3.13$^{+0.58}_{-0.41}$ & 0.19$^{+0.07}_{-0.05}$\\
3 & 0.4 && 2.0 & 2.0 && 3.10$^{+0.60}_{-0.41}$ & 0.42$^{+0.15}_{-0.12}$\\
4 & 0.05 && 4.8 & 1.7 && 4.31$^{+1.52}_{-0.79}$ & 0.05$^{+0.02}_{-0.01}$\\
4 & 0.15 && 2.7 & 1.7 && 4.39$^{+1.47}_{-0.85}$ & 0.15$^{+0.05}_{-0.04}$\\
4 & 0.2 && 2.4 & 1.7 && 4.38$^{+1.69}_{-0.85}$ & 0.19$^{+0.08}_{-0.06}$\\
4 & 0.4 && 1.7 & 1.7 && 4.38$^{+1.55}_{-0.86}$ & 0.38$^{+0.14}_{-0.11}$\\
10  &  0.5   &    &1.1 &1.2 && 11.54$^{+4.58}_{-3.88}$& 0.51$^{+0.16}_{-0.11}$\\
10  &  1.0   &    &0.8 &1.2 && 11.55$^{+4.69}_{-3.81}$& 1.02$^{+0.29}_{-0.22}$\\
10  &  1.5   &    &0.6 &1.2 && 12.29$^{+4.32}_{-3.89}$& 1.45$^{+0.29}_{-0.28}$
\enddata 
\end{deluxetable}

\begin{deluxetable}{cccccccc} 
\tablecolumns{8} 
\tablewidth{0pc} 
\tablecaption{Inversion of Numerical Data From Simulations\label{table:numerical}} 
\tablehead{ 
\multicolumn{2}{c}{Simulation Parameters}  & \colhead{}& \multicolumn{2}{c}{Fitted Data}  &\colhead{} & \multicolumn{2}{c}{Inversion Results} \\ 
\cline{1-2}\cline{4-5}\cline{7-8} \\ 
\colhead{$\zeta$} & \colhead{$l/R$}  & \colhead{} & \colhead{$L_{\rm g}/\lambda$}  &\colhead{$h/\lambda$}    & \colhead{} & \colhead{$\zeta$}    & \colhead{$l/R$}}
\startdata 
1.5 & 0.05 && 11.5 & 3.8 && 1.73$^{+0.12}_{-0.09}$ & 0.05$^{+0.02}_{-0.01}$\\
1.5 & 0.15 && 7.9 & 4.6 && 1.56$^{+0.08}_{-0.07}$ & 0.15$^{+0.05}_{-0.04}$\\
1.5 & 0.2 && 7.0 & 4.8 && 1.53$^{+0.08}_{-0.06}$ & 0.21$^{+0.07}_{-0.05}$\\
1.5 & 0.4 && 5.0 & 4.9 && 1.52$^{+0.07}_{-0.06}$ & 0.39$^{+0.09}_{-0.08}$\\

3 & 0.05 && 5.5 & 2.1 && 2.88$^{+0.46}_{-0.33}$ & 0.06$^{+0.02}_{-0.02}$\\
3 & 0.15 && 3.5 & 2.2 && 2.74$^{+0.44}_{-0.32}$ & 0.16$^{+0.06}_{-0.04}$\\
3 & 0.2 && 3.1 & 2.2 && 2.74$^{+0.41}_{-0.30}$ & 0.21$^{+0.07}_{-0.05}$\\
3 & 0.4 && 2.1 & 2.0 && 3.09$^{+0.57}_{-0.40}$ & 0.38$^{+0.13}_{-0.11}$\\

4 & 0.05 && 4.9 & 1.7 && 4.17$^{+1.32}_{-0.74}$ & 0.05$^{+0.02}_{-0.01}$\\
4 & 0.15 && 3.1 & 1.9 && 3.19$^{+0.64}_{-0.42}$ & 0.16$^{+0.06}_{-0.05}$\\
4 & 0.2 && 2.7 & 1.9 && 3.33$^{+0.74}_{-0.43}$ & 0.21$^{+0.07}_{-0.06}$\\
4 & 0.4 && 2.3 & 2.2 && 2.73$^{+0.43}_{-0.29}$ & 0.38$^{+0.12}_{-0.10}$\\

\enddata 
\end{deluxetable}

Then, simulations of transverse kink wave propagation in a magnetic flux tube were performed using a numerical code \citep[see][for details]{pascoe13}. A Lax-Wendroff code is used to solve the linear MHD equations in cylindrical coordinates. The lower boundary is driven harmonically with velocity perturbations corresponding to the loop footprint moving back and forth about its equilibrium. 
The simulation ends after 10 periods of oscillation and the spatial damping profile is investigated by considering the radial velocity component, $v_{\rm r}$ as a function of $z$ at the centre of the loop (Figure~\ref{dampingprofile}).  From the behavior of the amplitude of the excited kink waves at different heights the damping profile was fitted and values for $L_{\rm g}$ and $h$ obtained.  Using those fitted values as data, we repeated the inversion procedure. For the sake of comparison, parameter spaces that overlap with those in Table~\ref{table:synthetic} were considered. 

Figure~\ref{fig:posteriors} displays an example of the marginal posterior distributions and the joint probability distribution for $\zeta$ and $l/R$. For both parameters, well defined probability distributions are obtained. For each parameter, the median of the marginal posterior and errors given at the 68\%  credible interval are used to compute the estimates given in Table~\ref{table:numerical}. This Table shows the values for the physical parameters used in the simulations, the fitted length  scales, and the inferred physical parameters. Numerical and analytical forward models give similar results. This issue is discussed in detail by \cite{pascoe13} (see their figures 8, 9, and 10). Our Bayesian inference technique properly returns the physical parameters of interest. This means that the algebraic expressions for the forward problem are accurate enough, since their use correctly returns the values for the physical parameters that were actually used in the numerical setup.  As with synthetic data in Table~\ref{table:synthetic}, large density contrast values tend to produce larger errors in their determination by inversion.  The main problem lies in obtaining the parameters $L_{\rm g}$, $L_{\rm d}$, and $h$ from the data, and specifically in determining $h$ accurately, which determines the accuracy of the density estimate. 

The general spatial damping profile remains an accurate description of the damping behavior \citep{pascoe13}, for large density contrasts. The weak link is the errors in the least squares fit of the damping profile to the data. The 
errors are larger for higher density contrasts due to the Gaussian stage being very short, and the results are more sensitive to the errors in $h$. The lesson is that large contrasts are a challenge from the observational point of view, if the theoretical predictions by \cite{hood13,pascoe13} and the inversion technique presented here are to be employed.


\section{Conclusion}

The cross-field density structuring of magnetic waveguides cannot be currently estimated using seismology techniques. Yet such an understanding is crucial to assess the possible role of MHD waves in plasma heating processes. Recent theoretical results for the spatial damping of propagating MHD kink waves predict the existence of two different damping regimes for the spatial dependence of the velocity amplitude along the waveguides. In this Letter, we have explained how the observational identification of those regimes and the measurement of the associated damping length scales can be used to fully constrain the cross-field density structuring of magnetic waveguides.

Such measurements are currently unavailable and observational efforts towards the observational confirmation of the theoretical prediction  and the measurement of the associated spatial scales should be devised. They would not only provide us with hitherto undetermined information on density, but  would strongly endorse resonant absorption as damping/heating mechanism, since the presence of two damping regimes is an inherent property of this mechanism.
 
With the use of Bayesian inference, such measurements would ensure the inverse problem is consistently solved taking into account all the relevant information and with correct propagation of uncertainty. The range of parameters used in our inversion experiments covers the reasonable range of expected values used in previous studies \citep{goossens02a,ruderman02, arregui07a,goossens12b}, so the measurement and inversion of observations seem feasible.

The application to real observations would require high quality data, so that a reliable fitting can be performed to obtain the damping length scales and to be able to clearly discriminate between Gaussian and exponential damping regimes. As pointed out by \cite{pascoe13}, the Gaussian damping stage only applies for one wavelength, in the limit of density contrast tending to infinity. Fitting to such a small signal can lead to significant error bars. The method of fitting both damping regimes is applicable for low density contrast waveguides. Our application is limited to inference for a single oscillating magnetic tube. Observations by \cite{tomczyk07}  and \cite{tomczyk09} show that wave dynamics is ubiquitous over large coronal regions.  Our method could be extended in order to apply the inversion technique to such extended regions, thus obtaining overall information about the cross-field density structuring of the corona where these waves propagate. 

Recent studies have considered effects on the wave amplitude of propagating transverse kink waves 
from the consideration of a background field-aligned flow \citep{soler11d} and of longitudinal density stratification \citep{soler11c}. These two effects affect the wave amplitude and might partially or totally compensate the damping by resonant absorption, thus producing significant differences in the inferred parameters obtained in this study. Future seismology schemes should include these ingredients in their inversion process.

\acknowledgments
I.A. and A.A.R. acknowledge support by Ram\'on y Cajal Fellowships by the Spanish Ministry of Economy and Competitiveness (MINECO). I.A. acknowledges the support by the Spanish MICINN/MINECO and FEDER funds through project AYA2011-22846. A.A.R. acknowledges support by the Spanish MINECO through project AYA2010-18029 (Solar Magnetism and Astrophysical Spectropolarimetry) and from the Consolider-Ingenio 2010 CSD2009-00038 project. The computational work for this paper was carried out on the joint STFC and SFC (SRIF) funded cluster at the University of St Andrews (Scotland, UK).

\end{document}